\documentclass[12pt,reqno]{amsart}
\usepackage{graphicx}
\usepackage{amscd}
\usepackage{amsmath}
\usepackage{epsfig}
\usepackage{amsfonts}
\usepackage{amssymb}

\providecommand{\U}[1]{\protect\rule{.1in}{.1in}}
\providecommand{\U}[1]{\protect\rule{.1in}{.1in}}
\allowdisplaybreaks

\theoremstyle{plain}

\newtheorem{remark}{Remark}

\numberwithin{equation}{section}

\begin{document}
\title[Soliton-like solutions]{Soliton-like solutions for nonlinear Schr\"{o}%
dinger equation with variable quadratic Hamiltonians}
\author{Erwin Suazo}
\address{Department of Mathematical Sciences, University of Puerto Rico,
Mayaquez, call box 9000, PR 00681--9000, Puerto Rico}
\email{erwin.suazo@upr.edu}
\author{Sergei K. Suslov}
\address{School of Mathematical and Statistical Sciences \& Mathematical, Computational and Modeling Sciences Center, Arizona
State University, Tempe, AZ 85287--1804, U.S.A.}
\email{suslov@math.asu.edu}
\urladdr{http://hahn.la.asu.edu/\symbol{126}suslov/index.html}
\date{\today }
\subjclass{Primary 35Q55, 35Q51. Secondary 81Q05.}
\keywords{Nonlinear Schr\"{o}dinger equation, Gross--Pitaevskii equation,
Bose--Einstein condensation, Feshbach resonance, fiber optics, generalized
harmonic oscillators, soliton-like solutions, Jacobian elliptic functions,
Painlev\'{e}~II transcendents.}

\begin{abstract}
We construct one soliton solutions for the nonlinear Schr\"{o}dinger
equation with variable quadratic Hamiltonians in a unified form by taking
advantage of a complete (super) integrability of generalized harmonic
oscillators. The soliton wave evolution in external fields with variable
quadratic potentials is totally determined by the linear problem, like
motion of a classical particle with acceleration, and the (self-similar)
soliton shape is due to a subtle balance between the linear Hamiltonian
(dispersion and potential) and nonlinearity in the Schr\"{o}dinger equation
by the standards of soliton theory. Most linear (hypergeometric, Bessel) and
a few nonlinear (Jacobian elliptic, second Painlev\'{e} transcendental)
classical special functions of mathematical physics are linked together
through these solutions, thus providing a variety of nonlinear integrable
cases. Examples include bright and dark solitons, and Jacobi elliptic and
second Painlev\'{e} transcendental solutions for several variable
Hamiltonians that are important for current research in nonlinear optics and
Bose--Einstein condensation. The Feshbach resonance matter wave soliton
management is briefly discussed from this new perspective.
\end{abstract}

\maketitle

\section{Introduction}

Advances of the past decades in nonlinear optics, Bose--Einstein
condensates, propagation of soliton waves in plasma physics and in other
fields of nonlinear science have involved a detailed study of nonlinear Schr%
\"{o}dinger equations (see, for example, \cite{Khawetal02}, \cite%
{Bruga:Sci10}, \cite{KudryashovBook10}, \cite{ZYan:Konotop09}, \cite%
{Zakh:Shab71} and references therein). In the theory of Bose--Einstein
condensation \cite{Dal:Giorg:Pitaevski:Str99}, \cite{Pit:StrinBook}, from a
general point of view, the dynamics of gases of cooled atoms in a magnetic
trap at very low temperatures can be described by an effective equation for
the condensate wave function known as the Gross--Pitaevskii (or nonlinear
Schr\"{o}dinger) equation \cite{BongsSengs04}, \cite{Kagan:Surkov:Shlyap96},
\cite{Kagan:Surkov:Shlyap97}, \cite{Kivsh:Alex:Tur01}, \cite{Oraevsky01} and
\cite{Per-G:Tor:Mont}. Experimental observations of dark and bright solitons
\cite{BurgeretalShlyap99}, \cite{Cataletal01}, \cite{Denschletal00}, \cite%
{Khayketal02} and bright soliton trains \cite{Khawetal02}, \cite%
{Strecker:etal02}, \cite{Srteckeretal03} in the presence of harmonic
confinement have generated considerable research interest in this area \cite%
{BongsSengs04}, \cite{Frantz10}.

The propagation of an optical pulse in a real fiber is also well described
by a nonlinear Schr\"{o}dinger equation for the envelope of wave functions
travelling inside the fiber \cite{AblowPrinTrub04}, \cite{Agrawal}, \cite%
{Bruga:Sci10}, \cite{Degas10}, \cite{Hasegawa}, \cite{KivishLuth-Dav98}. A
class of self-similar solutions that exists for physically realistic
dispersion and nonlinearity profiles in a fiber with anomalous group
velocity dispersion is discussed in \cite{Kruglovetal03}, \cite{Krugloveta05}%
, \cite{Moores96}, \cite{Moores01}, \cite{PonomAgr07}, \cite%
{Serkin:Hasrgawa00}, \cite{Serkinetal04}, which suggests, among other
things, a method of pulse compression and a model of steady-state
asynchronous laser mode locking \cite{Moores01}. Solutions of a
nonhomogeneous Schr\"{o}dinger equation are also known for propagation of
soliton waves in plasma physics \cite{Balakrish85}, \cite{ChenHH:LiuCS76},
\cite{ChenHH:LiuCS78}, \cite{Newell78}.

Integration techniques of the nonlinear Schr\"{o}dinger equation include
Painlev\'{e} analysis \cite{AlKha10}, \cite{Bruga:Sci10}, \cite{Conte89},
\cite{Conte99}, \cite{ConteFordyPick93}, \cite{ConteMusette09}, \cite%
{Heetal09}, \cite{KudryashovBook10}, \cite{MusConte03}, \cite{Weissetal82},
Hirota method \cite{Hirota71}, \cite{HirotaBook}, \cite{KudryashovBook10},
Lax method \cite{AlKha10}, \cite{KudryashovBook10}, \cite{Lax68}, \cite%
{Zakh:Shab71}, Miura transformation \cite{Miura68}, \cite{Miuraetal68},
inverse scattering transform and Hamiltonian approach \cite{AblowClark91},
\cite{Ablowitzetal73}, \cite{Ablo:Seg81}, \cite{Fadd:Takh}, \cite%
{Gardneretal67}, \cite{Novikovetal} among others \cite{ChenHH74}, \cite%
{DegasRes}, \cite{DubMatvNov81}, \cite{Mateev:SalleBook}, \cite{OlverPBook},
\cite{Rogers:SchiefBook}. Although the classical soliton concept was
developed for nonlinear autonomous dispersive systems with time being an
independent variable only, not appearing in the nonlinear evolution
equations (see \cite{Serkinetal07}, \cite{Serkinetal10} for highlighting
this point), connections between autonomous and nonautonomous Schr\"{o}%
dinger equations have been discussed in \cite{AblowClark91}, \cite%
{AblowPrinTrub04}, \cite{Clark88}, \cite{HeLi10}, \cite{Kundu09}, \cite{Moores01}, \cite%
{Per-GTorrKonot06}, \cite{PonomAgr07} and \cite{Zhuk99} (see Remark~2 for an
explicit transformation). The formation of matter wave solitons in
Bose--Einstein condensation by magnetically tuning the interatomic
interaction near the Feshbach resonance provides an example of nonautonomous
systems that are currently under investigation \cite{BongsSengs04}, \cite%
{Frantz10}, \cite{Srteckeretal03}.

We elaborate on results of recent papers \cite{Khawetal02}, \cite{AlKha10},
\cite{Atre:Pani:Aga06}, \cite{Bruga:Sci10}, \cite{Eba:Khal}, \cite%
{LiangZhangLiu05}, \cite{ChenYi05}, \cite{HeLi10}, \cite{Heetal09}, \cite%
{Kruglovetal03}, \cite{Krugloveta05}, \cite{Kundu09}, \cite{Kudryashov09},
\cite{Roz}, \cite{Serkin:Hasrgawa00}, \cite{Serkin:Hasegawa00}, \cite%
{Serkinetal04}, \cite{Serkinetal07}, \cite{Serkinetal10}, \cite%
{Strecker:etal02}, \cite{Trall-Gin:Drake:Lop-Rich:Trall-Herr:Bir}, \cite%
{ZYan03a}, \cite{Zyan04}, \cite{ZYan10}, \cite{ZYan:Konotop09} on
construction of exact solutions of the nonlinear Schr\"{o}dinger equation
with variable quadratic Hamiltonians (see also \cite{Tao09} and \cite%
{Zakh:Shab71}, \cite{ZakhShab74}, \cite{ZakhShab79}). In this paper, a
unified form of these soliton-like (self-similar) solutions is presented,
thus combining progress of the soliton theory with a complete integrability
of generalized harmonic oscillators. We show, in general, that the soliton
evolution in external fields described by variable quadratic potentials is
totally determined by the linear problem, similar to the motion of a
classical particle with acceleration, while the original soliton shape is
due to a delicate balance between the linear Hamiltonian (dispersion and
potential) and nonlinearity in the Schr\"{o}dinger equation according to
basic principles of the soliton theory. Examples include bright and dark
solitons, and Jacobi elliptic and Painlev\'{e}~II transcendental solutions
for solitary wave profiles, which are important in nonlinear optics \cite%
{AblowPrinTrub04}, \cite{ChenHH:LiuCS76}, \cite{ChenHH:LiuCS78}, \cite%
{Kruglovetal03}, \cite{Krugloveta05}, \cite{Moores96}, \cite%
{Serkin:Hasrgawa00}, \cite{Serkinetal04}, \cite{Zakh:Shab71} and
Bose--Einstein condensation \cite{Atre:Pani:Aga06}, \cite{Khawetal02}, \cite%
{Serkinetal10}, \cite{Strecker:etal02}, \cite{ZYan10}.

The paper is organized as follows. We present a unified form of one soliton
solutions with integrability conditions, and sketch the proof in the next
two sections, respectively. In Section~4, more details are provided and some
simple examples are discussed. Section~5 deals with a Feshbach resonance
management of matter wave solitons. In the last section, an extension of our
method is given and a classical example of accelerating soliton in a
linearly inhomogeneous plasma \cite{ChenHH:LiuCS76}, \cite{ChenHH:LiuCS78}
is revisited from a new perspective. An attempt to collect most relevant
bibliography is made but in view of a rich history \footnote{%
Ref.~\cite{DegasRes} presents a detailed source on classical papers in the
soliton theory.} and the very high publication rate in these research areas
we must apologize in advance if some important papers are missing.

\section{Soliton-Like Solutions}

The nonlinear Schr\"{o}dinger equation%
\begin{equation}
i\frac{\partial \psi }{\partial t}=H\psi +g\psi +h\left\vert \psi
\right\vert ^{2}\psi ,  \label{s1}
\end{equation}%
where the variable Hamiltonian $H$ is a quadratic form of operators $%
p=-i\partial /\partial x$ and $x,$ namely,%
\begin{equation}
i\psi _{t}=-a\left( t\right) \psi _{xx}+b\left( t\right) x^{2}\psi -ic\left(
t\right) x\psi _{x}-id\left( t\right) \psi +g\left( x,t\right) \psi +h\left(
t\right) \left\vert \psi \right\vert ^{2}\psi  \label{s2}
\end{equation}%
($a,$ $b,$ $c,$ $d$ are suitable real-valued functions of time only) has the
following soliton-like solutions%
\begin{eqnarray}
\psi \left( x,t\right) &=&\frac{e^{i\phi }}{\sqrt{\mu \left( t\right) }}\exp
\left( i\left( \alpha \left( t\right) x^{2}+\beta \left( t\right) xy+\gamma
\left( t\right) y^{2}\right) \right)  \label{s3} \\
&&\times F\left( \beta \left( t\right) x+2\gamma \left( t\right) y\right)
\notag
\end{eqnarray}%
($\phi $ is a real constant, $y$ is a parameter and $\mu ,$ $\alpha ,$ $%
\beta ,$ $\gamma $ are real-valued functions of time only given by equations
(\ref{MKernel})--(\ref{C0}) below), provided that%
\begin{equation}
g=g_{0}a\left( t\right) \beta ^{2}\left( t\right) \left( \beta \left(
t\right) x+2\gamma \left( t\right) y\right) ^{m},\qquad h=h_{0}a\left(
t\right) \beta ^{2}\left( t\right) \mu \left( t\right)  \label{s4}
\end{equation}%
($g_{0}$ and $h_{0}$ are constants and $m=0,1).$ As we shall see in the next
section, these (integrability) conditions control the balance between the linear Hamiltonian
(dispersion and potential) and nonlinearity in the Schr\"{o}dinger equation (%
\ref{s2}) thus making possible an existence of the soliton-like solution
(with damping or amplification) in the presence of variable quadratic
potentials (see also \cite{Atre:Pani:Aga06}, \cite{Heetal09}, \cite%
{Serkinetal07}, \cite{Serkinetal10} and \cite{Zhangetal08} for discussion of
important special cases; Remark~2 provides an important interpretation of
relations (\ref{s4}) as a complete integrability condition for the
nonautonomous nonlinear Schr\"{o}dinger equation (\ref{s1}) when $m=0).$

Here, the soliton profile function $F\left( z\right) $ of a single
travelling wave-type argument $z=$ $\beta x+2\gamma y$ satisfies the
ordinary nonlinear differential equation of the form%
\begin{equation}
F^{\prime \prime }\left( z\right) =g_{0}z^{m}F\left( z\right)
+h_{0}F^{3}\left( z\right) .  \label{s5}
\end{equation}%
If $m=0,$ with the help of an integrating factor,%
\begin{equation}
\left( \frac{dF}{dz}\right) ^{2}=C_{0}+g_{0}F^{2}+\frac{1}{2}%
h_{0}F^{4}\qquad \left( C_{0}\text{ is a constant}\right) ,  \label{s5a}
\end{equation}%
which can be solved in terms of Jacobian elliptic functions \cite%
{AkhiezerElliptic}, \cite{Erd}, \cite{KudryashovBook10}, \cite{Whi:Wat}.
When $m=1,$ equation (\ref{s5}) leads to Painlev\'{e}~II transcendents \cite%
{Ablo:Seg81}, \cite{Clark10}, \cite{Conte99}, \cite{ConteMusette09}, \cite%
{KudryashovBook10}.

The variable phase is given in terms of solutions of the following system of
ordinary differential equations:%
\begin{equation}
\frac{d\alpha }{dt}+b+2c\alpha +4a\alpha ^{2}=0,  \label{s6}
\end{equation}%
\begin{equation}
\frac{d\beta }{dt}+\left( c+4a\alpha \right) \beta =0,  \label{s7}
\end{equation}%
\begin{equation}
\frac{d\gamma }{dt}+a\beta ^{2}=0  \label{s8}
\end{equation}%
(see Ref.~\cite{Cor-Sot:Lop:Sua:Sus} and the next section for more details),
where the standard substitution%
\begin{equation}
\alpha =\frac{1}{4a\left( t\right) }\frac{\mu ^{\prime }\left( t\right) }{%
\mu \left( t\right) }-\frac{d\left( t\right) }{2a\left( t\right) }
\label{s9}
\end{equation}%
reduces the Riccati equation (\ref{s6}) to the second order linear equation%
\begin{equation}
\mu ^{\prime \prime }-\tau \left( t\right) \mu ^{\prime }+4\sigma \left(
t\right) \mu =0  \label{s10}
\end{equation}%
with%
\begin{equation}
\tau \left( t\right) =\frac{a^{\prime }}{a}-2c+4d,\qquad \sigma \left(
t\right) =ab-cd+d^{2}+\frac{d}{2}\left( \frac{a^{\prime }}{a}-\frac{%
d^{\prime }}{d}\right) .  \label{s11}
\end{equation}%
(Relations with the corresponding Ehrenfest theorem for the linear
Hamiltonian are discussed in Ref.~\cite{Cor-Sot:Sua:SusInv}.)

It is worth noting that in the soliton-like solution under consideration (%
\ref{s3}) linear and nonlinear factors are essentially separated, namely,
the nonlinear part is represented only by the profile function $F$ of a
single travelling wave variable $z=\beta x+2\gamma y$ as solution of the
nonlinear equation (\ref{s5}). Letting $\beta x+2\gamma y=$constant, one
obtains%
\begin{equation}
x^{\prime }+\frac{\beta ^{\prime }}{\beta }x=2a\beta y  \label{s11a}
\end{equation}%
and%
\begin{equation}
x^{\prime \prime }-\frac{a^{\prime }}{a}x^{\prime }+\left( \left( \frac{%
\beta ^{\prime }}{\beta }\right) ^{\prime }-\frac{a^{\prime }}{a}\frac{\beta
^{\prime }}{\beta }-\left( \frac{\beta ^{\prime }}{\beta }\right)
^{2}\right) x=0  \label{s11b}
\end{equation}%
for the soliton velocity and acceleration with the aid of (\ref{s8}). Then,
by (\ref{s6})--(\ref{s7}):%
\begin{equation}
x^{\prime \prime }-\frac{a^{\prime }}{a}x^{\prime }+\left( 4ab-c^{2}+c\left(
\frac{a^{\prime }}{a}-\frac{c^{\prime }}{c}\right) \right) x=0  \label{s11c}
\end{equation}%
that is similar to equation of motion of a classical particle (damped parametric
oscillations).

The initial value problem for the system (\ref{s6})--(\ref{s8}), which
corresponds to the linear Schr\"{o}dinger equation with a variable quadratic
Hamiltonian (generalized harmonic oscillators \cite{Berry85}, \cite%
{Dod:Mal:Man75}, \cite{Hannay85}, \cite{Wolf81}, \cite%
{Yeon:Lee:Um:George:Pandey93}), can be explicitly solved in terms of
solutions of our characteristic equation (\ref{s10}) as follows \cite%
{Cor-Sot:Lop:Sua:Sus}, \cite{Cor-Sot:Sua:SusInv}, \cite{Suaz:Sus}, \cite%
{Suslov10}:%
\begin{eqnarray}
&&\mu \left( t\right) =2\mu \left( 0\right) \mu _{0}\left( t\right) \left(
\alpha \left( 0\right) +\gamma _{0}\left( t\right) \right) ,  \label{MKernel}
\\
&&\alpha \left( t\right) =\alpha _{0}\left( t\right) -\frac{\beta
_{0}^{2}\left( t\right) }{4\left( \alpha \left( 0\right) +\gamma _{0}\left(
t\right) \right) },  \label{AKernel} \\
&&\beta \left( t\right) =-\frac{\beta \left( 0\right) \beta _{0}\left(
t\right) }{2\left( \alpha \left( 0\right) +\gamma _{0}\left( t\right)
\right) }=\frac{\beta \left( 0\right) \mu \left( 0\right) }{\mu \left(
t\right) }\lambda \left( t\right) ,  \label{BKernel} \\
&&\gamma \left( t\right) =\gamma \left( 0\right) -\frac{\beta ^{2}\left(
0\right) }{4\left( \alpha \left( 0\right) +\gamma _{0}\left( t\right)
\right) },  \label{CKernel}
\end{eqnarray}%
where%
\begin{eqnarray}
&&\alpha _{0}\left( t\right) =\frac{1}{4a\left( t\right) }\frac{\mu
_{0}^{\prime }\left( t\right) }{\mu _{0}\left( t\right) }-\frac{d\left(
t\right) }{2a\left( t\right) },  \label{A0} \\
&&\beta _{0}\left( t\right) =-\frac{\lambda \left( t\right) }{\mu _{0}\left(
t\right) },\qquad \lambda \left( t\right) =\exp \left( -\int_{0}^{t}\left(
c\left( s\right) -2d\left( s\right) \right) \ ds\right) ,  \label{B0} \\
&&\gamma _{0}\left( t\right) =\frac{1}{2\mu _{1}\left( 0\right) }\frac{\mu
_{1}\left( t\right) }{\mu _{0}\left( t\right) }+\frac{d\left( 0\right) }{%
2a\left( 0\right) }  \label{C0}
\end{eqnarray}%
provided that $\mu _{0}$ and $\mu _{1}$ are the standard solutions of
equation (\ref{s10}) corresponding to the following initial conditions $\mu
_{0}\left( 0\right) =0,$ $\mu _{0}^{\prime }\left( 0\right) =2a\left(
0\right) \neq 0$ and $\mu _{1}\left( 0\right) \neq 0,$ $\mu _{1}^{\prime
}\left( 0\right) =0.$ (Formulas (\ref{A0})--(\ref{C0}) correspond to Green's
function of generalized harmonic oscillators; see, for example, \cite%
{Cor-Sot:Lop:Sua:Sus}, \cite{Cor-Sot:Sua:SusInv}, \cite{Dodonov:Man'koFIAN87}%
, \cite{Malkin:Man'ko79}, \cite{Suaz:Sus}, \cite{Suslov10} and references
therein for more details.)

The continuity with respect to initial data,%
\begin{equation}
\lim_{t\rightarrow 0^{+}}\alpha \left( t\right) =\alpha \left( 0\right)
,\quad \lim_{t\rightarrow 0^{+}}\beta \left( t\right) =\beta \left( 0\right)
,\quad \lim_{t\rightarrow 0^{+}}\gamma \left( t\right) =\gamma \left(
0\right) ,  \label{cont}
\end{equation}%
has been established in \cite{Suaz:Sus} for suitable smooth coefficients of
the linear Schr\"{o}dinger equation. Thus the soliton-like solution (\ref{s3}%
) evolves to the future $t>0$ starting from the following initial data:%
\begin{eqnarray}
\psi \left( x,0\right) &=&\lim_{t\rightarrow 0^{+}}\psi \left( x,t\right)
\label{s12} \\
&=&\frac{e^{i\phi }}{\sqrt{\mu \left( 0\right) }}\exp \left( i\left( \alpha
\left( 0\right) x^{2}+\beta \left( 0\right) xy+\gamma \left( 0\right)
y^{2}\right) \right)  \notag \\
&&\times F\left( \beta \left( 0\right) x+2\gamma \left( 0\right) y\right) ,
\notag
\end{eqnarray}%
where $\phi ,$ $\mu \left( 0\right) ,$ $\alpha \left( 0\right) ,$ $\beta
\left( 0\right) ,$ $\gamma \left( 0\right) $ and $y$ are arbitrary real
parameters (see also (\ref{ap12h}) for a more general solution of this form).

\begin{remark}
When $m=0,$ the gauge transformation $\psi =\chi \left( x,t\right) \exp %
\left[ ig_{0}\left( \gamma \left( t\right) -\gamma \left( 0\right) \right) %
\right] $ changes the original equation (\ref{s2}) into%
\begin{equation}
i\chi _{t}=-a\left( t\right) \chi _{xx}+b\left( t\right) x^{2}\chi -ic\left(
t\right) x\chi _{x}-id\left( t\right) \chi +h\left( t\right) \left\vert \chi
\right\vert ^{2}\chi ,  \label{s2a}
\end{equation}%
where $a,$ $b,$ $c,$ $d$ are suitable real-valued functions of time $t$ only
and%
\begin{equation}
h\left( t\right) =h_{0}\beta ^{2}\left( 0\right) \mu ^{2}\left( 0\right)
\frac{a\left( t\right) \lambda ^{2}\left( t\right) }{\mu \left( t\right) },
\label{s2b}
\end{equation}%
which is more common in practice. Once again, classical solution of the
linear equation (\ref{s10}), namely, our characteristic function $\mu \left(
t\right) ,$ completely controls the specific form of the nonlinearity factor
$h\left( t\right) $ required for creation of the soliton (an extension is
given in Section~6; see also Refs.~\cite{Atre:Pani:Aga06}, \cite%
{Serkinetal07} and \cite{Serkinetal10} for important special cases).
\end{remark}

\begin{remark}
A simple change of variables,%
\begin{equation}
\psi =\frac{1}{\sqrt{\mu }}e^{i\alpha x^{2}}\chi \left( \beta x,\gamma
\right) ,  \label{s2c}
\end{equation}%
transforms the nonautonomous equation (\ref{s2}) with conditions (\ref{s4}),
when $m=0,$ into a standard autonomous nonlinear Schr\"{o}dinger equation
with respect to the new variables $\xi =\beta x$ and $\tau =\gamma :$%
\begin{equation}
i\chi _{t}+g_{0}\chi +h_{0}\left\vert \chi \right\vert ^{2}\chi =\chi _{\xi
\xi },  \label{s3d}
\end{equation}%
which is completely integrable by advanced methods of the soliton theory
\cite{AblowClark91}, \cite{Ablo:Seg81}, \cite{KudryashovBook10}, \cite%
{Zakh:Shab71} (see also \cite{DegasRes} and references cited in the
introduction). This observation provides an alternative approach to
derivation of our equations (\ref{s6})--(\ref{s9}). An extension of the
transformation (\ref{s2c}) is given in \cite{Suslov11}.
\end{remark}

\section{Sketch of the Proof}

Following \cite{Cor-Sot:Lop:Sua:Sus} (see also \cite{ChenHH:LiuCS76}, \cite%
{Merz} and \cite{Roz}), we are looking for exact solutions of the form%
\begin{equation}
\psi =A\left( x,t\right) e^{iS\left( x,t\right) },\quad S\left( x,t\right)
=\alpha \left( t\right) x^{2}+\beta \left( t\right) xy+\gamma \left(
t\right) y^{2}  \label{P1}
\end{equation}%
($y$ is a parameter). Substituting into (\ref{s2}) and taking the imaginary
part,%
\begin{equation}
A_{t}+\left( \left( 4a\alpha +c\right) x+2a\beta y\right) A_{x}+\left(
2\alpha a+d\right) A=0.  \label{P2}
\end{equation}%
For the real part, equating coefficients of all admissible powers of $%
x^{m}y^{n}$ with $m+n=2,$ one gets our system of ordinary differential
equations (\ref{s6})--(\ref{s8}) of the corresponding linear Schr\"{o}dinger
equation with the unique solution (\ref{MKernel})--(\ref{C0}) already
obtained in Refs.~\cite{Cor-Sot:Lop:Sua:Sus}, \cite{Suaz:Sus}, \cite%
{Suslov10} and/or elsewhere. In addition, an auxiliary nonlinear equation of
the form%
\begin{equation}
aA_{xx}=gA+hA^{3}  \label{P3}
\end{equation}%
appears as a contribution from the last two terms. With the help of (\ref{s7}%
) and (\ref{s9}) our equation (\ref{P2}) can be rewritten as%
\begin{equation}
A_{t}-\left( \frac{\beta ^{\prime }}{\beta }x-2a\beta y\right) A_{x}+\frac{1%
}{2}\frac{\mu ^{\prime }}{\mu }A=0.  \label{P4}
\end{equation}%
Looking for a travelling wave solution with damping or amplification:
\begin{equation}
A=A\left( x,t\right) =\frac{1}{\sqrt{\mu \left( t\right) }}\ F\left(
z\right) ,\qquad z=c_{0}\left( t\right) x+c_{1}\left( t\right) y,  \label{P5}
\end{equation}%
one gets%
\begin{equation}
c_{0}^{\prime }x+c_{1}^{\prime }y=\left( \frac{\beta ^{\prime }}{\beta }%
x-2a\beta y\right) c_{0}  \label{P6}
\end{equation}%
with $c_{0}=\beta $ and $c_{1}=2\gamma $ (or $z=\beta x+2\gamma y).$ Then
equation (\ref{P3}) takes the form%
\begin{equation}
\frac{d^{2}}{dz^{2}}F\left( z\right) =\frac{g}{a\beta ^{2}}F\left( z\right) +%
\frac{h}{a\beta ^{2}\mu }F^{3}\left( z\right) ,  \label{P7}
\end{equation}%
which must have all coefficients depending on $z$ only in order to preserve
a self-similar profile of the travelling wave with damping or amplification.
This results in the required equation (\ref{s5}) under the balancing
conditions (\ref{s4}) and our proof is complete. (An extension is given in
Section~6.)

\begin{remark}
Assuming in (\ref{P4}) that%
\begin{equation}
A\left( x,t\right) =\frac{1}{\sqrt{\mu \left( t\right) }}\ B\left( \xi ,\tau
\right) ,  \label{P8}
\end{equation}
where $\xi =\beta x$ and $\tau =2\gamma y,$ implies $B_{\xi }=B_{\tau }$
with a general solution $B\left( \xi ,\tau \right) =F\left( \xi +\tau
\right) =F\left( \beta x+2\gamma y\right) .$
\end{remark}

\section{Details and Examples}

A brief description of the method under consideration is as follows. In
order to obtain soliton-like solutions (\ref{s3}) explicitly, say in terms
of elementary and/or transcendental functions, one has to solve, in general,
the nonlinear equation (\ref{s5}) for the profile function $F\left( z\right)
$ in terms of Jacobian elliptic functions \cite{AkhiezerElliptic}, \cite{Erd}%
, \cite{KudryashovBook10}, \cite{Rain}, \cite{Whi:Wat} (some elementary
solutions are also available), when $m=0,$ or in terms of Painlev\'{e}~II
transcendents, when $m=1$ (it is known that if $m>1,$ this equation does not
have the Painlev\'{e} property \cite{Ablo:Seg81}, \cite{KudryashovBook10}).
In addition, one has to solve the linear characteristic equation (\ref{s10}%
), which has a variety of solutions in terms of elementary and special
(hypergeometric, Bessel) functions \cite{An:As:Ro}, \cite{Mag:Win}, \cite%
{Ni:Uv}, \cite{Rain}, \cite{Wa}. Many elementary solutions of the
corresponding linear Schr\"{o}dinger equation for generalized harmonic
oscillators are known explicitly (see, for example, \cite%
{Cor-Sot:Lop:Sua:Sus}, \cite{Cor-Sot:Sua:Sus}, \cite{Cor-Sot:Sua:SusInv},
\cite{Cor-Sot:Sus}, \cite{Dodonov:Man'koFIAN87}, \cite{Malkin:Man'ko79},
\cite{Suslov10}, \cite{Wolf81}, \cite{Yeon:Lee:Um:George:Pandey93} and
references therein). Then, the linear part allows determination of the
travelling wave argument $z=\beta x+2\gamma y$ and the damping (or
amplifying) factor $\mu ^{-1/2}$ of the soliton-like solution (\ref{s3}).
Our balancing conditions (\ref{s4}) control dispersion, potential and
nonlinearity in the original nonlinear Schr\"{o}dinger equation (\ref{s2}),
which is crucial for the soliton existence. (An extension is discussed in
Section~6.)

\subsection{Nonlinear Part}

When $m=0,$ equation (\ref{s5}) is integrated to the first order equation (%
\ref{s5a}) and (the corresponding initial value problem) can be solved by
the reduction of elliptic integrals in terms of Jacobian or Weierstrass (doubly)
periodic elliptic functions \cite{AkhiezerElliptic}, \cite{Erd}, \cite%
{Krugloveta05}, \cite{KudryashovBook10}. We are interested in real-valued
solutions. Some of the classical nonlinear wave configurations are given by%
\begin{eqnarray}
F\left( z\right) &=&\left( \frac{g_{0}+\sqrt{g_{0}^{2}-2C_{0}h_{0}}}{-h_{0}}%
\right) ^{1/2}  \label{cn} \\
&&\times \text{cn}\left( \left( g_{0}^{2}-2C_{0}h_{0}\right) ^{1/4}z,\left(
\frac{g_{0}+\sqrt{g_{0}^{2}-2C_{0}h_{0}}}{2\sqrt{g_{0}^{2}-2C_{0}h_{0}}}%
\right) ^{1/2}\right) ,  \notag
\end{eqnarray}%
if $h_{0}<0$ and%
\begin{eqnarray}
F\left( z\right) &=&\left( \frac{-g_{0}+\sqrt{g_{0}^{2}-2C_{0}h_{0}}}{h_{0}}%
\right) ^{1/2}  \label{sn} \\
&&\times \text{sn}\left( \left( \frac{C_{0}h_{0}}{-g_{0}+\sqrt{%
g_{0}^{2}-2C_{0}h_{0}}}\right) z,\left( \frac{g_{0}-\sqrt{%
g_{0}^{2}-2C_{0}h_{0}}}{g_{0}+\sqrt{g_{0}^{2}-2C_{0}h_{0}}}\right)
^{1/2}\right) ,  \notag
\end{eqnarray}%
if $g_{0}<0.$ Here, cn$\left( u,k\right) $ and sn$\left( u,k\right) $ are
the Jacobi elliptic functions \cite{AkhiezerElliptic}, \cite{Erd}, \cite%
{Whi:Wat}. Familiar special cases include the \textit{bright} soliton:%
\begin{equation}
F\left( z\right) =\sqrt{\frac{2g_{0}}{-h_{0}}}\frac{1}{\cosh \left( \sqrt{%
g_{0}}z\right) }  \label{cosh}
\end{equation}%
with $C_{0}=0$ in (\ref{cn}) and the \textit{dark} soliton:%
\begin{equation}
F\left( z\right) =\sqrt{\frac{-g_{0}}{h_{0}}}\tanh \left( \sqrt{\frac{-g_{0}%
}{2}}z\right)  \label{tanh}
\end{equation}%
with $C_{0}=g_{0}^{2}/\left( 2h_{0}\right) $ in (\ref{sn}), when cn$\left(
u,1\right) =1/\cosh u$ and sn$\left( u,1\right) =\tanh u,$ respectively (the
real period tends to infinity). More details can be found in Refs.~\cite%
{AkhiezerElliptic}, \cite{Erd}, \cite{FedMurShlyap99}, \cite{Krugloveta05},
\cite{Whi:Wat}, \cite{Zakh:Shab71} and/or elsewhere.

If $m=1,$ the substitution $F\left( z\right) =g_{0}^{1/3}\sqrt{2/h_{0}}\
w\left( \zeta \right) $ and $\zeta =zg_{0}^{1/3}$ transforms (\ref{s5}) into
the second Painlev\'{e} equation,%
\begin{equation}
w^{\prime \prime }=\zeta w+2w^{3}.  \label{PainII}
\end{equation}%
In the limit $w\rightarrow 0,$ this equation reduces to the Airy equation
and its solution may be thought of as a nonlinear generalization of an Airy
function \cite{AbloSeg77}. There is a one-parameter family of real solutions
$w=A_{k}\left( \zeta \right) $ that are bounded for all real $\zeta $ with
the following asymptotic properties:%
\begin{equation}
A_{k}\left( \zeta \right) =\left\{
\begin{array}{c}
k\text{Ai\/}\left( \zeta \right) ,\qquad \zeta \rightarrow +\infty \bigskip
\\
r\left\vert \zeta \right\vert ^{-1/4}\sin \left( s\left( \zeta \right)
-\theta _{0}\right) +\text{o}\left( \left\vert \zeta \right\vert
^{-1/4}\right) ,\quad \zeta \rightarrow -\infty \bigskip%
\end{array}%
\right.  \label{AsPII}
\end{equation}%
Here, Ai\/$\left( \zeta \right) $ is the Ai\/ry function, $-1<k<1$ provided $%
k\neq 0,$ $r^{2}=-\pi ^{-1}\ln \left( 1-k^{2}\right) ,$%
\begin{equation}
s\left( \zeta \right) =\frac{2}{3}\left\vert \zeta \right\vert ^{3/2}-\frac{3%
}{4}r^{2}\ln \left\vert \zeta \right\vert  \label{AsPIIs}
\end{equation}%
and%
\begin{equation}
\theta _{0}=\frac{3}{2}r^{2}\ln 2+\arg \Gamma \left( 1-\frac{i}{2}%
r^{2}\right) +\frac{\pi }{4}\left( 1-2\text{sign\/}\left( k\right) \right) .
\label{AsPIIph}
\end{equation}%
These asymptotics were found in \cite{AbloSeg77}, \cite{SegurAb81} and,
eventually, had been proven rigorously in \cite{DeiftZhou93}, \cite%
{DeiftZhou95} (see \cite{Ablo:Seg81}, \cite{Bassometal98}, \cite{Clark10},
\cite{ClarkMcLeod88}, \cite{Conte99}, \cite{ConteMusette09}, \cite{Takei02}
and references therein for study of this nonlinear Airy function; graphs of
these functions are presented in \cite{Clark10}, see Figure~1 for an
example: $w_{.5}=A_{1/2}\left( x\right) .)$ An application to a soliton
moving with a constant velocity in linearly inhomogeneous plasma is
discussed in Section~6.
%
%\FRAME{ftbpFU}{5.499in}{3.1938in}{0pt}{\Qcb{Graphs of
%the Airy function $.5$Ai\/$\left( x\right) $ and the nonlinear Airy function
%$w_{.5}=A_{1/2}\left( x\right) $ (red and blue in an electronic version,
%respectively, \protect\cite{Clark10}).}}{}{asolfig.png}{\special{language
%"Scientific Word";type "GRAPHIC";maintain-aspect-ratio TRUE;display
%"USEDEF";valid_file "F";width 5.499in;height 3.1938in;depth
%0pt;original-width 8.6225in;original-height 4.9898in;cropleft "0";croptop
%"1";cropright "1";cropbottom "0";filename 'ASolFig.png';file-properties
%"XNPEU";}}
%
%
%%%%%%%%%%%%%%%%%%%%%%%%%%%%%%%%%%%%%%%%%%%
%%%This WinEdt version of the figure~1%%%
\begin{figure}[ptbh]
\centering\scalebox{.65}{\includegraphics{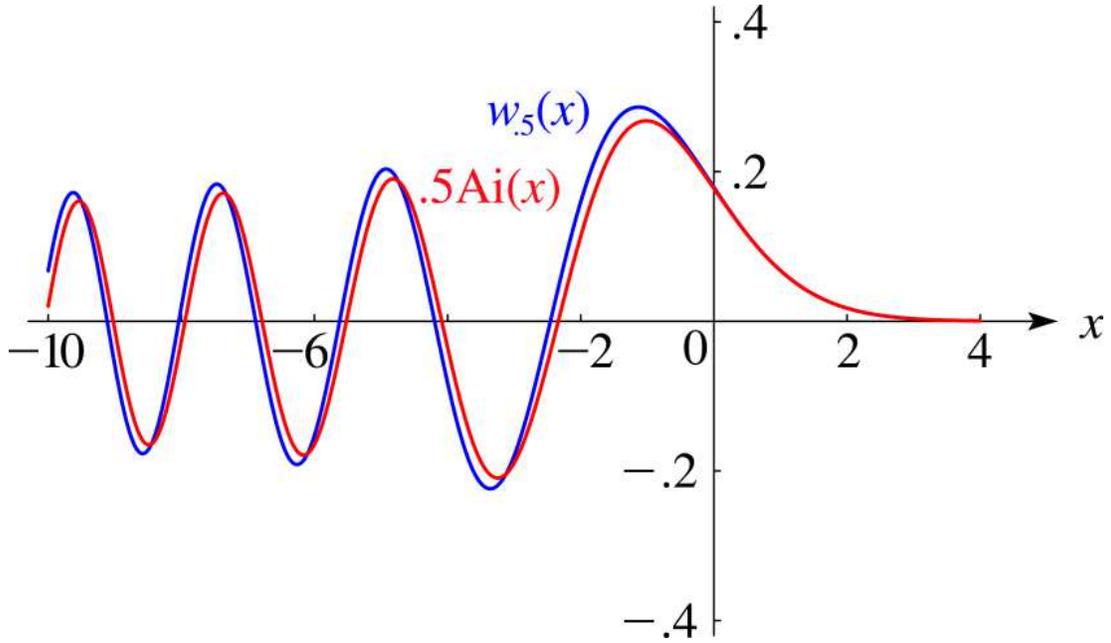}}
\caption{Graphs of the Airy function $.5$Ai\/$\left( x\right) $ and the nonlinear Airy function
$w_{.5}=A_{1/2}\left( x\right)$ (red and blue in an electronic version, respectively, \protect\cite{Clark10}).}
\end{figure}
%%%%%%%%%%%%%%%%%%%%%%%%%%%%%%%%%%%%%%%%%%%
%
%

\subsection{Linear Part}

Generalized harmonic oscillators \cite{Berry85}, \cite{Dod:Mal:Man75}, \cite%
{Hannay85}, \cite{Wolf81}, \cite{Yeon:Lee:Um:George:Pandey93}, which
correspond to the Schr\"{o}dinger equation with variable quadratic
Hamiltonians, are well studied in quantum mechanics (see also \cite%
{Cor-Sot:Lop:Sua:Sus}, \cite{Cor-Sot:Sua:Sus}, \cite{Dodonov:Man'koFIAN87},
\cite{Lan:Sus}, \cite{Malkin:Man'ko79}, \cite{Suslov10} and references
therein for a general approach and known elementary and transcendental
solutions). A few examples include the Caldirola--Kanai Hamiltonian of the
quantum damped oscillator \cite{Caldirola41}, \cite{Dekker81}, \cite{Kanai48}%
, \cite{Um:Yeon:George} and some of its natural modifications; a modified
oscillator considered by Meiler, Cordero-Soto and Suslov \cite{Me:Co:Su},
\cite{Cor-Sot:Sus}, and the degenerate parametric oscillator \cite%
{Cor-Sot:SusDPO}; the quantum damped oscillator of Chru\'{s}ci\'{n}ski and
Jurkowski \cite{Chru:Jurk}, and a quantum-modified parametric oscillator
among others. Green's functions are derived in a united way in Ref.~\cite%
{Cor-Sot:Sua:SusInv}.

\subsection{Examples}

Combination of linear and nonlinear parts by our formula (\ref{s3}) results
in numerous explicit soliton-like solutions for corresponding nonlinear Schr%
\"{o}dinger equations. It is worth noting that in this approach most linear
and some nonlinear classical special functions of mathematical physics are
linked together through these solutions.

\subsubsection{Nonlinear Optics}

In the simplest form,%
\begin{equation}
i\frac{\partial \psi }{\partial t}=\frac{\partial ^{2}\psi }{\partial x^{2}}%
+g\psi +h\left\vert \psi \right\vert ^{2}\psi ,  \label{ex1}
\end{equation}%
one gets \cite{Cor-Sot:Lop:Sua:Sus}, \cite{Moores96}, \cite{Moores01}%
\begin{eqnarray}
&&\alpha \left( t\right) =\frac{\alpha _{0}}{1-4\alpha _{0}t},\qquad \qquad
\beta \left( t\right) =\frac{\beta _{0}}{1-4\alpha _{0}t},  \label{ex2} \\
&&\gamma \left( t\right) =\gamma _{0}+\frac{\beta _{0}^{2}t}{1-4\alpha _{0}t}%
,\quad \mu \left( t\right) =\mu _{0}\left( 1-4\alpha _{0}t\right)  \notag
\end{eqnarray}%
($\mu _{0},$ $\alpha _{0},$ $\beta _{0},$ $\gamma _{0}$ are constants) and%
\begin{eqnarray}
&&z=\frac{\beta _{0}x+2\left( \gamma _{0}+\left( \beta _{0}^{2}-4\alpha
_{0}\gamma _{0}\right) t\right) y}{1-4\alpha _{0}t},  \label{ex3} \\
&&g\left( x,t\right) =-\frac{g_{0}\beta _{0}^{2}}{\left( 1-4\alpha
_{0}t\right) ^{2}}z^{m}\qquad \left( m=0,1\right) ,  \label{ex4} \\
&&h\left( t\right) =-\frac{h_{0}\mu _{0}\beta _{0}^{2}}{1-4\alpha _{0}t}.
\label{ex5}
\end{eqnarray}%
(Traditionally, $\alpha _{0}=0$ and $m=0$ with $\psi =\chi \exp \left(
ig_{0}\beta _{0}^{2}t\right) $ \cite{KudryashovBook10}, \cite{Zakh:Shab71}.
The case $m=1$ is discussed in Section~6.)

The case $b=c=0,$%
\begin{equation}
i\psi _{t}=-a\psi _{xx}-id\psi +g\psi +h\left\vert \psi \right\vert ^{2}\psi
,  \label{ex6}
\end{equation}%
is of interest in fiber optics (see, for example, \cite{AblowPrinTrub04}, \cite{Agrawal}, \cite%
{Hasegawa}, \cite{KivishLuth-Dav98}, \cite{Kruglovetal03}, \cite%
{Krugloveta05}, \cite{Moores96}, \cite{Moores01}, \cite{PonomAgr07}, \cite%
{Rajuetal05}, \cite{Serkin:Hasrgawa00}, \cite{Serkin:Hasegawa00}, \cite%
{Serkinetal07}, \cite{Serkinetal10} and references therein). Here, all
parameters $a\left( t\right) ,$ $d\left( t\right) $ and $h\left( t\right) $
are functions of the propagation distance $t=z$ and this equation describes
the amplification or attenuation (if $d$ is positive) of pulses propagating
nonlinearly in a single-mode optical fiber, where $\psi \left( t,x\right) $
is the complex envelope of the electrical field in a comoving frame, $x=\tau
$ is the retarded time, $a\left( t\right) $ is the group velocity dispersion
parameter, $d\left( t\right) $ is the dispersion gain or loss function, and $%
h\left( t\right) $ is the nonlinearity parameter \cite{Kruglovetal03}, \cite%
{Krugloveta05}.

The substitution $\psi =\chi e^{-\Lambda },$ $\Lambda \left( t\right)
=\int_{0}^{t}d\left( s\right) \ ds$ results in%
\begin{equation}
i\frac{\partial \chi }{\partial t}=-a\frac{\partial ^{2}\chi }{\partial x^{2}%
}+g\chi +he^{-2\Lambda }\left\vert \chi \right\vert ^{2}\chi ,  \label{ex7}
\end{equation}%
which, of course, can be solved by the method under consideration, but a
standard change of the time variable,%
\begin{equation}
\tau =-\int_{0}^{t}a\left( s\right) \ ds,  \label{ex8}
\end{equation}%
transforms this equation into the previous one. Just replace $t\rightarrow
\tau $ (see also \cite{Kruglovetal03}, \cite{Krugloveta05}, \cite%
{Serkin:Hasrgawa00}, \cite{Serkin:Hasegawa00} and \cite{Serkinetal04}, where
this simple observation has been omitted). More general transformations are
discussed in Refs.~\cite{AblowClark91}, \cite{Clark88}, \cite{Cor-Sot:Sua:Sus}, \cite{Kundu09}, \cite{Moores01},
\cite{Per-GTorrKonot06}, \cite{PonomAgr07}, \cite{Suslov11}, \cite{Zhuk99}.

\subsubsection{Harmonic Solitons}

In a similar fashion, one can show that the nonlinear Schr\"{o}dinger
equation of the form%
\begin{equation}
i\frac{\partial \chi }{\partial t}=\frac{1}{2}\left( -\frac{\partial
^{2}\chi }{\partial x^{2}}+x^{2}\chi \right) +\frac{h_{0}\mu _{0}\beta
_{0}^{2}}{2\left( \cos t+2\alpha _{0}\sin t\right) }\left\vert \chi
\right\vert ^{2}\chi  \label{ex8a}
\end{equation}%
has the following explicit solution:%
\begin{equation}
\chi \left( x,t\right) =\frac{e^{iS\left( x,t\right) }}{\sqrt{\left\vert \mu
_{0}\right\vert \left( \cos t+2\alpha _{0}\sin t\right) }}\ F\left( z\right)
,  \label{ex8b}
\end{equation}%
where%
\begin{equation}
z=\frac{\beta _{0}x+\left( 2\gamma _{0}\cos t-\left( \beta _{0}^{2}-4\alpha
_{0}\gamma _{0}\right) \sin t\right) y}{\cos t+2\alpha _{0}\sin t}
\label{ex8c}
\end{equation}%
and%
\begin{eqnarray}
S\left( x,t\right) &=&\frac{2\alpha _{0}\cos t-\sin t}{2\left( \cos
t+2\alpha _{0}\sin t\right) }x^{2}+\frac{\beta _{0}xy}{\cos t+2\alpha
_{0}\sin t}  \label{ex8d} \\
&&+\frac{2\gamma _{0}\cos t-\left( \beta _{0}^{2}-4\alpha _{0}\gamma
_{0}\right) \sin t}{2\left( \cos t+2\alpha _{0}\sin t\right) }y^{2}+\frac{%
g_{0}\beta _{0}^{2}\sin t}{2\left( \cos t+2\alpha _{0}\sin t\right) }  \notag
\end{eqnarray}%
provided%
\begin{equation}
F^{\prime \prime }=g_{0}F+h_{0}F^{3}  \label{ex8e}
\end{equation}%
($m=0,$ $\mu _{0}\neq 0,$ $\alpha _{0},$ $\beta _{0},$ $\gamma _{0},$ $%
g_{0}, $ $h_{0}$ and $y$ are arbitrary real constants). The reader may
choose the profile function $F,$ say in one of the forms (\ref{cn})--(\ref%
{tanh}).

If $m=1,$ the Schr\"{o}dinger equation,%
\begin{eqnarray}
i\frac{\partial \psi }{\partial t} &=&\frac{1}{2}\left( -\frac{\partial
^{2}\psi }{\partial x^{2}}+x^{2}\psi \right) +\frac{g_{0}\beta _{0}^{2}}{%
2\left( \cos t+2\alpha _{0}\sin t\right) ^{2}}z\psi  \label{HarmASol1} \\
&&+\frac{h_{0}\mu _{0}\beta _{0}^{2}}{2\left( \cos t+2\alpha _{0}\sin
t\right) }\left\vert \psi \right\vert ^{2}\psi ,  \notag
\end{eqnarray}%
has a solution-like solution of the form (\ref{ex8b}) (last term in (\ref%
{ex8d}) should be omitted) provided that%
\begin{equation}
F^{\prime \prime }=g_{0}zF+h_{0}F^{3},  \label{HarmASol2}
\end{equation}%
which solution is given in terms of the nonlinear Airy function $A_{k}\left(
\zeta \right) .$ Further details are left to the reader (see also Section~6).

\section{Matter Wave Solitons}

\subsection{Gross--Pitaevskii equation}

Discovery of Bose--Einstein condensates in ultra-cold gases of weakly
interacting alkali-metal atoms has stimulated intensive studies of nonlinear
matter waves on a macroscopic scale (see, for example, \cite{BongsSengs04},
\cite{CornWieNobel}, \cite{Frantz10}, \cite{KetterleNobel}, \cite%
{Srteckeretal03}). The Gross--Pitaevskii equation for a zero-temperature
condensate of atoms, confined in a cylindrical trap $V_{0}\left( x,y\right)
=m\omega _{\bot }^{2}\left( x^{2}+y^{2}\right) /2,$ and a time-dependent
harmonic confinement, which can be either attractive or expulsive, along the
$z$ direction, $V_{1}\left( z,t\right) =m\omega _{0}^{2}\left( t\right)
z^{2}/2,$ is given by \cite{Atre:Pani:Aga06}, \cite%
{Dal:Giorg:Pitaevski:Str99}, \cite{Erdetal07}, \cite{LiebSeiringYngv00},
\cite{Pit:StrinBook}, \cite{Salasnetal02}, \cite{Serkinetal10}:%
\begin{equation}
i\hslash \frac{\partial \Psi \left( \boldsymbol{r},t\right) }{\partial t}%
=\left( -\frac{\hslash ^{2}}{2m}\Delta +V_{\text{ext}}\left( \boldsymbol{r}%
,t\right) +U\left\vert \Psi \left( \boldsymbol{r},t\right) \right\vert ^{2}+i%
\frac{\hslash \eta \left( t\right) }{2}\right) \Psi \left( \boldsymbol{r}%
,t\right) ,  \label{GPEq}
\end{equation}%
where $U=4\pi \hslash ^{2}a_{s}/m,$ $a_{s}$ is the $s$-wave scattering
length, $m$ is the mass of the atom, $V_{\text{ext}}=V_{0}\left( x,y\right)
+V_{1}\left( z,t\right) $ and the condensate interaction with the normal
atomic cloud through three-body interaction is phenomenologically
incorporated by a gain or loss term $\eta \left( t\right) .$ If the
interaction energy of atoms is much less that the kinetic energy in the
transverse direction, then the substitution
\begin{eqnarray}
&&\Psi \left( \boldsymbol{r},t\right) =\frac{1}{\sqrt{2\pi a_{0}}a_{\bot }}%
\exp \left( -i\omega _{\bot }t-\frac{x^{2}+y^{2}}{2a_{\bot }^{2}}+\frac{%
\Lambda \left( t\right) }{2}\right)  \label{ex9} \\
&&\quad \qquad \quad \times \psi \left( \frac{z}{a_{\bot }},\omega _{\bot
}t\right) ,\qquad \qquad \Lambda \left( t\right) =\int_{0}^{t}\eta \left(
s\right) \ ds  \notag
\end{eqnarray}%
allows one to reduce the three-dimensional Gross--Pitaevskii equation (\ref%
{GPEq}) to the following one-dimensional nonlinear Schr\"{o}dinger equation
in new dimensionless units $\zeta =z/a_{\bot }$ and $\tau =\omega _{\bot }t:$%
\begin{equation}
i\frac{\partial \psi }{\partial \tau }=\frac{1}{2}\left( -\frac{\partial
^{2}\psi }{\partial \zeta ^{2}}+\omega ^{2}\left( \tau \right) \zeta
^{2}\psi \right) +\kappa \left( \tau \right) \left\vert \psi \right\vert
^{2}\psi .  \label{ex10}
\end{equation}%
Here,%
\begin{equation}
\kappa \left( \tau \right) =2e^{\Lambda }\frac{a_{s}}{a_{0}},\qquad \omega
^{2}\left( \tau \right) =\frac{\omega _{0}^{2}}{\omega _{\bot }^{2}},\qquad
a_{\bot }=\sqrt{\frac{\hslash }{m\omega _{\bot }}}  \label{ex10a}
\end{equation}%
and $a_{0}$ is the Bohr radius (see Refs.~\cite{JackKavPeth98}, \cite%
{Kivsh:Alex:Tur01}, \cite{MenString02}, \cite{MunDelg07}, \cite{MunDelg08},
\cite{Per-GMich98}, \cite{Salasnetal02} for more details).

Letting%
\begin{equation}
\psi =\chi \exp \left( i\int_{0}^{\tau }g\left( s\right) \ ds\right)
\label{ex11}
\end{equation}%
with the function $g\left( \tau \right) $ given by (\ref{Gg}) below,
equation (\ref{ex10}) can be transformed into (\ref{s2}), where%
\begin{equation}
a=\frac{1}{2},\quad b=\frac{1}{2}\omega ^{2}\left( \tau \right) ,\quad
c=d=0,\quad h=\kappa \left( \tau \right) .  \label{ex12}
\end{equation}%
Then our characteristic equation (\ref{s10}) take the form%
\begin{equation}
\mu ^{\prime \prime }+\omega ^{2}\left( \tau \right) \mu =0,
\label{ClassOsc}
\end{equation}%
which describes the motion of a classical oscillator with variable frequency
\cite{Mag:Win}. (This equation coincides also with the Ehrenfest theorem for
the corresponding linear Hamiltonian \cite{Cor-Sot:Sua:SusInv}.) Choosing
the standard solutions $\mu _{0}\left( \tau \right) $ and $\mu _{1}\left(
\tau \right) $ with $\mu _{0}\left( 0\right) =0,$ $\mu _{0}^{\prime }\left(
0\right) =1$ and $\mu _{1}\left( 0\right) \neq 0,$ $\mu _{1}^{\prime }\left(
0\right) =0,$ one can use formulas (\ref{MKernel})--(\ref{C0}) with $c=d=0$
in order to solve the linear problem in quadratures. This gives the soliton
travelling wave variable $z=\beta x+2\gamma y$ and the following balancing
conditions:%
\begin{eqnarray}
&&\kappa \left( \tau \right) =h_{0}\frac{\beta ^{2}\left( 0\right) \mu
^{2}\left( 0\right) }{2\mu \left( \tau \right) },  \label{Ep} \\
&&g\left( \tau \right) =g_{0}\frac{\beta ^{2}\left( 0\right) \mu ^{2}\left(
0\right) }{2\mu ^{2}\left( \tau \right) },\qquad \int_{0}^{\tau }g\left(
s\right) \ ds=g_{0}\left( \gamma \left( 0\right) -\gamma \left( \tau \right)
\right)  \label{Gg}
\end{eqnarray}%
when $m=0.$

\subsection{Feshbach Resonance}

The properties of Bose--Einstein condensed gases can be strongly altered by
tuning the external magnetic field. A Feshbach resonance management for
Bose--Einstein condensates has been discussed from experimental and
theoretical perspectives by many authors (see, for example, \cite%
{Abdullaevetal03}, \cite{ChinVogKett03}, \cite{Courteilletal98}, \cite%
{Cornishetal00}, \cite{CornWieNobel}, \cite{FedKagShlyapWal}, \cite{Heetal09}%
, \cite{Houbetal98}, \cite{Inouyetal98}, \cite{KaganSurShlyap97PRL}, \cite%
{Kevreketal03}, \cite{LiangZhangLiu05}, \cite{MatInfMalTrip05}, \cite%
{MoerVerhAxels95}, \cite{PelinKevrekFrantz03}, \cite{Per-GKonotBrazh04},
\cite{Pit:StrinBook}, \cite{Robertsetal98}, \cite{Serkinetal10}, \cite%
{Stengetal99}, \cite{Srteckeretal03}, \cite{Stwalley76}, \cite%
{TimmetalPhysRep99}, \cite{TiesVerhSt93}, \cite{Zhangetal08} and references
therein). The Feshbach resonance is a scattering resonance in which pairs of
free atoms are tuned via Zeeman effect into resonance with vibrational state
of the diatomic molecule \cite{CornWieNobel}, \cite{Srteckeretal03}, \cite%
{TiesVerhSt93}. (They are known as Feshbach resonance because of their
similarity to scattering resonances described by Herman Feshbach in nuclear
collisions.) The strength of the nonlinearity $U$ is defined in terms of $s$%
-wave scattering length $a_{s},$ namely,%
\begin{equation}
U=\frac{4\pi \hslash ^{2}a_{s}}{m},  \label{3DNonLinConst}
\end{equation}%
and dependence of atomic collision cross section due to existence of the
metastable state \cite{CornWieNobel}, \cite{FedKagShlyapWal}, \cite%
{KaganSurShlyap97PRL}, \cite{Stengetal99} enables $a_{s}$ to be continuously
tuned from positive to negative values. (The scattering length also
determines the formation rate, the spectrum of collective excitations, the
evolution of the condensate phase, the coupling with the noncondensed atoms,
and other important properties \cite{Cornishetal00}, \cite{Pit:StrinBook}.)
As follows from the experiments, the $s$-wave scattering length is the
following function of the applied magnetic field \cite{MoerVerhAxels95}:%
\begin{equation}
\frac{a_{s}\left( B\right) }{a_{0}}=a_{\infty }\left( 1+\frac{\Delta _{0}}{%
B_{0}-B}\right)  \label{FeshRes}
\end{equation}%
(The Feshbach resonance provides, so to speak, a continuous knob to adjust
the atom-atom interaction from repulsive to attractive, and from weak to
strong \cite{Srteckeretal03}. Thus it is possible to study strongly
interacting, weakly or noninteracting, or collapsing condensates \cite%
{KaganSurShlyap97PRL}, all with the same alkali species and experimental
setup. When the nonlinearity $U=0,$ one deals with linear modes of a
macroscopic harmonic oscillator \cite{Kivsh:Alex:Tur01}; see, for example,
\cite{Lan:Sus}, \cite{Dodonov:Man'koFIAN87}, \cite{Malkin:Man'ko79} and
references therein for a detailed treatment of the corresponding quantum
oscillator with variable frequency.) In the empirically established
expression (\ref{FeshRes}), $B_{0}$ is the resonant value of the magnetic
field, $a_{\infty }$ is the off-resonance scattering length and parameter $%
\Delta _{0}$ represents the resonance width in units of the Bohr radius $%
a_{0}$ (see \cite{CornWieNobel}, \cite{MoerVerhAxels95}, \cite{Serkinetal10}%
, \cite{TimmetalPhysRep99}, \cite{TiesVerhSt93} and references therein for
more details). Feshbach resonances have been observed in $^{85}Rb$ at $164~G$
\cite{Courteilletal98}, \cite{Robertsetal98}, \cite{Cornishetal00}, in $%
^{23}Na$ at $853$ and $907~G$ \cite{Inouyetal98} and have also been
identified in $^{6}Li$ \cite{Houbetal98}, \cite{O'Haraetal02}, \cite%
{Strecker:etal02}, \cite{Srteckeretal03}.

\subsection{Matter Wave Soliton Management}

The Feshbach resonance provides an effective practical tool for experimental
study of the matter wave solitons. Indeed, for creation of a certain soliton
configuration one needs to satisfy the following condition:
\begin{equation}
h_{0}\frac{\beta ^{2}\left( 0\right) \mu ^{2}\left( 0\right) }{4\mu }%
e^{-\Lambda }=a_{\infty }\left( 1+\frac{\Delta _{0}}{B_{0}-B}\right)
\label{BalFesh}
\end{equation}%
in order to synchronize the Feshbach resonance and harmonic trap. (Here,
both sides have the same simple pole structure, which can be used in
experimental setting.) This equation allows determination of the classical
law of motion (kinematics in $z$ direction) of the expectation value $\mu
=\langle \Psi ,\boldsymbol{r}\Psi \rangle $ with respect to the linear part
of Gross--Pitaevskii Hamiltonian (\ref{GPEq}), when $U=\eta =0,$ in terms of
\ a suitable applied magnetic field $B$ near the Feshbach resonance. (The
synchronized harmonic trap oscillation frequency should be found from the
classical equation of motion (\ref{ClassOsc}) as $\omega ^{2}=-\mu ^{\prime
\prime }/\mu .)$ Vice versa, the required tuning magnetic field is given by%
\begin{equation}
B=B_{0}+\frac{4a_{\infty }\Delta _{0}e^{\Lambda }\mu }{4a_{\infty }\Delta
_{0}e^{\Lambda }\mu -h_{0}\beta ^{2}\left( 0\right) \mu ^{2}\left( 0\right) }%
,  \label{MagFesh}
\end{equation}%
if a particular law of motion $\mu $ is obtain by integration (dynamics) of
the classical equation (\ref{ClassOsc}). Our criteria of the wave matter
soliton management are consistent with ones obtained in Refs.~\cite%
{Atre:Pani:Aga06}, \cite{Heetal09}, \cite{Serkinetal07}, \cite{Serkinetal10}
and \cite{Zhangetal08}, if the classical equation of motion (\ref{ClassOsc})
is taken into account (this point seems not emphasized in these papers).

\subsection{Examples}

Harmonic matter wave solitons, which correspond to $\omega ^{2}=\omega
_{0}^{2}/\omega _{\bot }^{2}=$constant in the nonlinear Schr\"{o}dinger
equation (\ref{ex10}), namely,%
\begin{equation}
i\frac{\partial \psi }{\partial \tau }=\frac{1}{2}\left( -\frac{\partial
^{2}\psi }{\partial \zeta ^{2}}+\omega ^{2}\zeta ^{2}\psi \right) +2\frac{%
a_{s}}{a_{0}}\left\vert \psi \right\vert ^{2}\psi ,\qquad \eta =0
\label{HarmSchrEq}
\end{equation}%
can be produced in Bose--Einstein condensates by tuning the external
magnetic field near the Feshbach resonance as follows%
\begin{equation}
B=B_{0}+\Delta _{0}+\frac{\omega h_{0}\mu _{0}\beta _{0}^{2}\Delta _{0}}{%
4a_{\infty }\left( 2\alpha _{0}\sin \omega \tau +\omega \cos \omega \tau
\right) -\omega h_{0}\mu _{0}\beta _{0}^{2}}.  \label{HarmFesh}
\end{equation}%
Letting $\omega \rightarrow 0,$ one gets%
\begin{equation}
B=B_{0}+\Delta _{0}+\frac{h_{0}\mu _{0}\beta _{0}^{2}\Delta _{0}}{4a_{\infty
}\left( 2\alpha _{0}\tau +1\right) -h_{0}\mu _{0}\beta _{0}^{2}},
\label{LinFesh}
\end{equation}%
when $\omega _{0}=0$ (a similar case has been recently discussed in Refs.~%
\cite{Heetal09} and \cite{Serkinetal10}; it is of interest to analyze
possible experimental setup; see also \cite{Zhangetal08}).

The reader may find more details on the synchronization of Feshbach
resonance and harmonic trap, explicit soliton configurations, and available
numerical and experimental results in recent papers \cite{Atre:Pani:Aga06},
\cite{Heetal09}, \cite{Serkinetal07}, \cite{Serkinetal10}, \cite{ZYan10}
(see also \cite{Khawetal02}, \cite{AlKha10}, \cite{Bruga:Sci10}, \cite%
{Eba:Khal}, \cite{Kruglovetal03}, \cite{Krugloveta05}, \cite%
{Trall-Gin:Drake:Lop-Rich:Trall-Herr:Bir}, \cite{ZYan:Konotop09}, \cite%
{Zakh:Shab71} and references therein).

\section{Generalization}

If an arbitrary linear combination of operators $p=-i\partial /\partial x$
and $x$ is added to the quadratic Hamiltonian in equation (\ref{s2}), namely,%
\begin{equation}
i\psi _{t}=-a\left( t\right) \psi _{xx}+b\left( t\right) x^{2}\psi -ic\left(
t\right) x\psi _{x}-id\left( t\right) \psi -f\left( t\right) x\psi +ig\left(
t\right) \psi _{x}+h\left( t\right) \left\vert \psi \right\vert ^{2}\psi ,
\label{ap1}
\end{equation}%
one can look for exact solutions in a more general form%
\begin{eqnarray}
\psi &=&A\left( x,t\right) e^{iS\left( x,t\right) },  \label{ap2} \\
S\left( x,t\right) &=&\alpha \left( t\right) x^{2}+\beta \left( t\right)
xy+\gamma \left( t\right) y^{2}+\delta \left( t\right) x+\varepsilon \left(
t\right) y+\kappa \left( t\right) +\xi \left( t\right)  \notag
\end{eqnarray}%
($y$ is a parameter, we are separating contributions from linear $\kappa $
and nonlinear $\xi $ parts in the constant term). The linear part has been
already solved in \cite{Cor-Sot:Lop:Sua:Sus} and \cite{Suaz:Sus}. One has
additional equations%
\begin{equation}
\frac{d\delta }{dt}+\left( c+4a\alpha \right) \delta =f+2\alpha g,
\label{ap3}
\end{equation}%
\begin{equation}
\frac{d\varepsilon }{dt}=\left( g-2a\delta \right) \beta ,  \label{ap4}
\end{equation}%
\begin{equation}
\frac{d\kappa }{dt}=g\delta -a\delta ^{2}  \label{ap5}
\end{equation}%
to the system (\ref{s6})--(\ref{s8}), whose solutions are given by%
\begin{eqnarray}
\delta \left( t\right) &=&\delta _{0}\left( t\right) -\frac{\beta _{0}\left(
t\right) \left( \delta \left( 0\right) +\varepsilon _{0}\left( t\right)
\right) }{2\left( \alpha \left( 0\right) +\gamma _{0}\left( t\right) \right)
},  \label{ap6} \\
\varepsilon \left( t\right) &=&\varepsilon \left( 0\right) -\frac{\beta
\left( 0\right) \left( \delta \left( 0\right) +\varepsilon _{0}\left(
t\right) \right) }{2\left( \alpha \left( 0\right) +\gamma _{0}\left(
t\right) \right) },  \label{ap7} \\
\kappa \left( t\right) &=&\kappa \left( 0\right) +\kappa _{0}\left( t\right)
-\frac{\left( \delta \left( 0\right) +\varepsilon _{0}\left( t\right)
\right) ^{2}}{4\left( \alpha \left( 0\right) +\gamma _{0}\left( t\right)
\right) }.  \label{ap8}
\end{eqnarray}%
Here,%
\begin{equation}
\delta _{0}\left( t\right) =\frac{\lambda \left( t\right) }{\mu _{0}\left(
t\right) }\int_{0}^{t}\left[ \left( f\left( s\right) -\frac{d\left( s\right)
}{a\left( s\right) }g\left( s\right) \right) \mu _{0}\left( s\right) +\frac{%
g\left( s\right) }{2a\left( s\right) }\mu _{0}^{\prime }\left( s\right) %
\right] \frac{ds}{\lambda \left( s\right) },  \label{ap9}
\end{equation}%
\begin{eqnarray}
\varepsilon _{0}\left( t\right) &=&-\frac{2a\left( t\right) \lambda \left(
t\right) }{\mu _{0}^{\prime }\left( t\right) }\delta _{0}\left( t\right)
+8\int_{0}^{t}\frac{a\left( s\right) \sigma \left( s\right) \lambda \left(
s\right) }{\left( \mu _{0}^{\prime }\left( s\right) \right) ^{2}}\left( \mu
_{0}\left( s\right) \delta _{0}\left( s\right) \right) \ ds  \label{ap10} \\
&&\quad +2\int_{0}^{t}\frac{a\left( s\right) \lambda \left( s\right) }{\mu
_{0}^{\prime }\left( s\right) }\left( f\left( s\right) -\frac{d\left(
s\right) }{a\left( s\right) }g\left( s\right) \right) \ ds,  \notag
\end{eqnarray}%
\begin{eqnarray}
\kappa _{0}\left( t\right) &=&\frac{a\left( t\right) \mu _{0}\left( t\right)
}{\mu _{0}^{\prime }\left( t\right) }\delta _{0}^{2}\left( t\right)
-4\int_{0}^{t}\frac{a\left( s\right) \sigma \left( s\right) }{\left( \mu
_{0}^{\prime }\left( s\right) \right) ^{2}}\left( \mu _{0}\left( s\right)
\delta _{0}\left( s\right) \right) ^{2}\ ds  \label{ap11} \\
&&\quad -2\int_{0}^{t}\frac{a\left( s\right) }{\mu _{0}^{\prime }\left(
s\right) }\left( \mu _{0}\left( s\right) \delta _{0}\left( s\right) \right)
\left( f\left( s\right) -\frac{d\left( s\right) }{a\left( s\right) }g\left(
s\right) \right) \ ds  \notag
\end{eqnarray}%
with $\delta _{0}\left( 0\right) =-\varepsilon _{0}\left( 0\right) =g\left(
0\right) /\left( 2a\left( 0\right) \right) $ and $\kappa _{0}\left( 0\right)
=0$ (see Refs.~\cite{Cor-Sot:Lop:Sua:Sus} and \cite{Suaz:Sus} for more
details).

Our equation (\ref{P2}) takes the form%
\begin{equation}
A_{t}+\left( \left( 4a\alpha +c\right) x+2a\beta y+2\delta a-g\right)
A_{x}+\left( 2\alpha a+d\right) A=0  \label{ap12}
\end{equation}%
and the nonlinear equation (\ref{P3}) becomes%
\begin{equation}
aA_{xx}=\frac{d\xi }{dt}A+hA^{3}.  \label{ap12a}
\end{equation}%
Once again, with the help of (\ref{s7}), (\ref{s9}) and (\ref{ap4}),
equation (\ref{ap12}) can be rewritten as%
\begin{equation}
A_{t}-\left( \frac{\beta ^{\prime }}{\beta }x-2a\beta y+\frac{\varepsilon
^{\prime }}{\beta }\right) A_{x}+\frac{1}{2}\frac{\mu ^{\prime }}{\mu }A=0
\label{ap12b}
\end{equation}%
and looking for a travelling wave solution of the form%
\begin{equation}
A=A\left( x,t\right) =\frac{1}{\sqrt{\mu \left( t\right) }}\ F\left(
z\right) ,\qquad z=c_{0}\left( t\right) x+c_{1}\left( t\right) y+c_{2}\left(
t\right) ,  \label{ap12c}
\end{equation}%
one gets%
\begin{equation}
c_{0}^{\prime }x+c_{1}^{\prime }y+c_{2}^{\prime }=\left( \frac{\beta
^{\prime }}{\beta }x-2a\beta y+\frac{\varepsilon ^{\prime }}{\beta }\right)
c_{0}  \label{ap12d}
\end{equation}%
with $c_{0}=\beta ,c_{1}=2\gamma $ and $c_{2}=\varepsilon $ (or $z=\beta
x+2\gamma y+\varepsilon ).$ Then%
\begin{equation}
\frac{d^{2}}{dz^{2}}F\left( z\right) =\frac{d\xi /dt}{a\beta ^{2}}F\left(
z\right) +\frac{h}{a\beta ^{2}\mu }F^{3}\left( z\right)  \label{ap12e}
\end{equation}%
and our balancing conditions are given by%
\begin{equation}
\frac{d\xi }{dt}=g_{0}a\left( t\right) \beta ^{2}\left( t\right) ,\qquad
h=h_{0}a\left( t\right) \beta ^{2}\left( t\right) \mu \left( t\right)
=h_{0}\beta ^{2}\left( 0\right) \mu ^{2}\left( 0\right) \frac{a\left(
t\right) \lambda ^{2}\left( t\right) }{\mu \left( t\right) }  \label{ap12f}
\end{equation}%
with $\xi =g_{0}\left( \gamma \left( 0\right) -\gamma \left( t\right)
\right) $ according to (\ref{s8}). For the soliton velocity,%
\begin{equation}
x^{\prime }+\frac{\beta ^{\prime }}{\beta }x=2a\beta y+2a\delta -g,
\label{ap12g}
\end{equation}%
thus extending (\ref{s11a}). Equation of motion is given by%
\begin{eqnarray}
&&x^{\prime \prime }-\frac{a^{\prime }}{a}x^{\prime }+\left(
4ab-c^{2}+c\left( \frac{a^{\prime }}{a}-\frac{c^{\prime }}{c}\right) \right)
x  \label{ap12ga} \\
&&\qquad =2af+\left( \frac{a^{\prime }}{a}-c\right) g-g^{\prime }  \notag
\end{eqnarray}%
as a nonhomogeneous generalization of (\ref{s11c}) (forced damped parametric
oscillators).

As a final result, our solitary wave solution has the form%
\begin{eqnarray}
\psi \left( x,t\right) &=&\frac{e^{i\phi }}{\sqrt{\mu }}\exp \left( i\left(
\alpha x^{2}+\beta xy+\gamma \left( y^{2}-g_{0}\right) +\delta x+\varepsilon
y+\kappa \right) \right)  \label{ap12h} \\
&&\times F\left( \beta x+2\gamma y+\varepsilon \right) ,  \notag
\end{eqnarray}%
where the elliptic function $F$ satisfies the nonlinear equation (\ref{s5})
with $m=0$ and $\phi ,$ $y,$ $g_{0}$ and $h_{0}$ are real parameters.
Time-dependent functions $\mu \left( t\right) ,$ $\alpha \left( t\right) ,$ $%
\beta \left( t\right) ,$ $\gamma \left( t\right) ,$ $\delta \left( t\right) $
and $\kappa \left( t\right) $ are given by our equations (\ref{MKernel})--(%
\ref{C0}) and (\ref{ap6})--(\ref{ap11}) (as in the corresponding solution of
the linear Schr\"{o}dinger equation \cite{Cor-Sot:Lop:Sua:Sus}, \cite%
{Suaz:Sus}).

\textit{Example~1}. A soliton motion with acceleration in linearly
inhomogeneous plasma was discovered in Refs.~\cite{ChenHH:LiuCS76} and \cite%
{ChenHH:LiuCS78} (see also \cite{Balakrish85}, \cite{TappZab71}). For a
modified equation,%
\begin{equation}
i\frac{\partial \psi }{\partial t}+\frac{\partial ^{2}\psi }{\partial x^{2}}%
+2kx\psi =\frac{h_{0}\mu _{0}\beta _{0}^{2}}{1+4\alpha _{0}t}\left\vert \psi
\right\vert ^{2}\psi ,  \label{ihp1}
\end{equation}%
where $k,$ $h_{0},$ $\alpha _{0},$ $\beta _{0}$ and $\mu _{0}$ are
constants, we get $\mu \left( t\right) =\mu _{0}\left( 1+4\alpha
_{0}t\right) $ and%
\begin{eqnarray}
&&\alpha \left( t\right) =\frac{\alpha _{0}}{1+4\alpha _{0}t},\qquad \qquad
\quad \beta \left( t\right) =\frac{\beta _{0}}{1+4\alpha _{0}t},
\label{ihp2} \\
&&\gamma \left( t\right) =\gamma _{0}-\frac{\beta _{0}^{2}t}{1+4\alpha _{0}t}%
,\quad \qquad \delta \left( t\right) =kt+\frac{\delta _{0}+kt}{1+4\alpha
_{0}t},  \notag \\
&&\varepsilon \left( t\right) =\varepsilon _{0}-\frac{2\beta _{0}t\left(
\delta _{0}+kt\right) }{1+4\alpha _{0}t},\quad \kappa \left( t\right)
=\kappa _{0}-\frac{k^{2}t^{3}}{3}-\frac{t\left( \delta _{0}+kt\right) ^{2}}{%
1+4\alpha _{0}t}  \notag
\end{eqnarray}%
with%
\begin{eqnarray}
z &=&\beta x+2\gamma y+\varepsilon  \label{ihp3} \\
&=&\beta _{0}\frac{x-2t\left( \beta _{0}y+\delta _{0}+kt\right) }{1+4\alpha
_{0}t}+2y\gamma _{0}+\varepsilon _{0}  \notag
\end{eqnarray}%
in our particular solution (\ref{ap12h}). The classical case \cite%
{ChenHH:LiuCS76}, \cite{ChenHH:LiuCS78} corresponds to $\alpha _{0}=0$ and $%
h_{0}\mu _{0}\beta _{0}^{2}=-2$ (with $k\rightarrow -k).$ The reader may
choose the profile $F$ in one of the forms (\ref{cn})--(\ref{tanh}).

\textit{Example~2}. A similar case occur, if one takes $m=1$ in (\ref{ex4}).
The corresponding Schr\"{o}dinger equation is given by%
\begin{equation}
i\frac{\partial \psi }{\partial t}+\frac{\partial ^{2}\psi }{\partial x^{2}}=%
\frac{g_{0}\beta _{0}^{2}}{\left( 1+4\alpha _{0}t\right) ^{2}}z\psi +\frac{%
h_{0}\mu _{0}\beta _{0}^{2}}{1+4\alpha _{0}t}\left\vert \psi \right\vert
^{2}\psi ,  \label{ihp4}
\end{equation}%
where%
\begin{equation}
z=\frac{\beta _{0}x+2\left( \gamma _{0}-\left( \beta _{0}^{2}-4\alpha
_{0}\gamma _{0}\right) t\right) y}{1+4\alpha _{0}t}.  \label{ihp5}
\end{equation}%
With the help of the gauge transformation,%
\begin{equation}
\psi =e^{-if\left( t\right) }\chi \left( x,t\right) ,\qquad \frac{df}{dt}%
=2g_{0}\beta _{0}^{2}y\frac{\gamma _{0}-\left( \beta _{0}^{2}-4\alpha
_{0}\gamma _{0}\right) t}{\left( 1+4\alpha _{0}t\right) ^{3}},  \label{ihp6}
\end{equation}%
one gets%
\begin{equation}
i\chi _{t}+\chi _{xx}-\frac{g_{0}\beta _{0}^{3}x}{\left( 1+4\alpha
_{0}t\right) ^{3}}\chi =\frac{h_{0}\mu _{0}\beta _{0}^{2}}{1+4\alpha _{0}t}%
\left\vert \chi \right\vert ^{2}\chi  \label{ihp7}
\end{equation}%
and%
\begin{equation}
\chi \left( x,t\right) =\frac{e^{iS\left( x,t\right) }}{\sqrt{\left\vert \mu
_{0}\right\vert \left( 1+4\alpha _{0}t\right) }}g_{0}^{1/3}\sqrt{\frac{2}{%
h_{0}}}A_{k}\left( g_{0}^{1/3}z\right) .  \label{ihp8}
\end{equation}%
Here,%
\begin{eqnarray}
S\left( x,t\right) &=&\frac{\alpha _{0}x^{2}+\beta _{0}xy+\left( \gamma
_{0}-\left( \beta _{0}^{2}-4\alpha _{0}\gamma _{0}\right) t\right) y^{2}}{%
1+4\alpha _{0}t}  \label{ihp9} \\
&&+g_{0}\beta _{0}^{2}t\frac{2\gamma _{0}-\left( \beta _{0}^{2}-8\alpha
_{0}\gamma _{0}\right) t}{1+4\alpha _{0}t}y  \notag
\end{eqnarray}%
and the soliton profile is defined (as a solution of the second Painlev\'{e}
equation) in terms of the nonlinear Airy function $A_{k}\left( \zeta \right)
$ with asymptotics given by (\ref{AsPII})--(\ref{AsPIIph}). (Graph of one of
these functions, $w_{.5}=A_{1/2},$ is presented on Figure~1 from \cite%
{Clark10}.) It is worth noting that, in a contrast to the previous case, our
$A$-soliton moves with a constant velocity when $\alpha _{0}=0.$ Further
details are left to the reader.

\noindent \textbf{Acknowledgements.\/} We thank Andrew Bremner, Carlos
Castillo-Ch\'{a}vez, Elliott ~H.~Lieb, Vladimir I.~Man'ko and Svetlana
Roudenko for support, valuable discussions and encouragement. One of us
(SKS) is grateful to his students in a Mathematics of Quantum Mechanics
course at Arizona State University for a detailed discussion of several
aspects of this work.

\end{document}